\documentstyle[12pt]{article}

\title{Degenerate Sectors of the Ashtekar Gravity}
\author{Jerzy Lewandowski and Jacek Wi\'sniewski}

\def\C{{\cal C}}
\def\G{{\cal G}}
\def\eq{\begin{equation}}
\def\eeq{\end{equation}}
\def\sl{sl(2,{\mbox {\boldmath $C$}})}

\begin{document}

\baselineskip=15pt
\maketitle

\newtheorem{lemat}{Lemma}
\newtheorem{fakt}{Fact} 

\begin{center}
 {\it Instytut Fizyki Teoretycznej, Wydzia\l \ Fizyki,\\
 Uniwersytet Warszawski, ul. Ho\.za 69, 00-681, Warszawa, Poland} 
\end{center}

\begin{abstract}
 This work completes the task of solving locally the Einstein-Ashtekar equations for degenerate data. The two remaining degenerate sectors of the classical $3+1$ dimensional theory are considered. First, with all densitized triad vectors linearly dependent and second, with only two independent ones. It is shown how to solve the Ashtekar-Einstein equations completely by suitable gauge fixing and choice of coordinates. Remarkably, the Hamiltonian weakly Poisson commutes with the conditions defining the sectors. The summary of degenerate solutions is given in the Appendix.
\end{abstract}

\section{Introduction}

\hspace{\parindent}

The standard Einstein's gravity theory corresponds to an
open region in the real section of the Ashtekar's theory phase space.  
The boundary\footnote{Meaning here just the closure of the region minus 
the region  itself.} of that region is set up by  degenerate data. 
There are several motivations to study the degenerate sector.
First, a natural question which arises is whether or not 
the evolution could throw some data out of the Einstein's 
theory region. But then, since the reality is preserved, 
the evolving data should cross the degenerate sector.
Second, according to the loop quantization, quantum excitations of  the 
gravitational field are lower  dimensional and define degenerate, 
non-invertible metric tensors. 

The degenerate data can be  classified with respect to the rank of the 
densitized triad, and the rank of the squared triad (see next section). 
It should be noted that all the considerations in this paper are local. 
Our classification of the degeneracy, in particular, applies
only to open regions of the surface of initial data,
whereas in a general case the types can vary from one region to 
another one.

All the solutions of the Einstein-Ashtekar equations
of the types (1,1) and (2,2) were derived in 
\cite{Jacobson,Lewandowski}.  In the first case \cite{Jacobson}, 
 a general solution is the congruence of the integral curves 
defined by the triad and foliating $\Sigma$ which behave like 1+1 
dimensional vacuum space-times with a pair of massless complex valued 
fields propagating along them. 
In the (2,2) case \cite{Lewandowski}, it was shown that the 
preservation of the reality by the evolution 
implies the existence of a foliation of $\Sigma$ into the integral
2-surfaces
tangent to a given triad. Analogously to the Jacobson's case, the
equations 
of the 3+1 gravity make the 2-surfaces behave like 2+1 dimensional empty 
space-times with an extra massless complex field assigned to each surface
and 
propagating along it. 
 An important observation was, that the conditions defining
each of the sectors Poisson commute  with the Hamiltonian
modulo themselves and  the constraints.
 
   In the present paper, the Einstein-Ashtekar equations 
will be solved for the remaining two types of the degenerate data.
In the first (1,0) case the solution is space-time which is a 
`set of independently evolving points'. In the second (2,1) case, 
the general solution is such that the surface of initial data $\Sigma$ 
is foliated by  integral curves of the vector field from the triad. 
Nine complex fields evolve along these curves.
As in the previously studied cases, it is  shown that
the conditions defining each degeneracy sector
weakly (in the same sense as above) Poisson commute with the 
Hamiltonian\footnote{Another interesting derivation of our result on the possibility of the evolution of a non-degenerate data into a degenerate one was given in \cite{Ma}.}.

Before the systematic study of the Ashtekar equations in the 
degenerate sector which was started by Jacobson \cite{Jacobson},
various aspects of the degenerate sector were discussed for instance by 
Jacobson and Romano \cite{Romano}, Bengtsson \cite{Bengtsson}, 
Reisenberger \cite{Reisenberger} and Matschull \cite{Matschull}. See also more recent work \cite{recent}. 

\section{Ashtekar's theory.}

\hspace{\parindent}
 For  reader's convenience we shall briefly review the Ashtekar's theory.

 It is a canonical theory on a space-time manifold $ \Sigma \times {\mbox {\boldmath $ R $}}$, where $\Sigma$ is a three-real-surface of initial data (the 'space') and {\boldmath $ R $} is the one dimensional space of values for a time parameter. The phase space consists of the pairs of fields $(A,E)$, where $A$ is an algebra $ sl(2,{\mbox {\boldmath $C$}}) $-valued one-form on $\Sigma$ and $E$ is an $ sl(2,{\mbox {\boldmath $C$}}) $-valued vector density field of weight 1 defined on $\Sigma$. Using local
 coordinates $ (x^{a})=(x^{1},x^{2},x^{3}) $ on $\Sigma$ and a basis $ (\tau_{i})=(\tau_{1},\tau_{2},\tau_{3}) $ of $ sl(2,{\mbox {\boldmath $C$}}) $ we write
\eq
 A= A^{i}_{a} \tau_{i} \otimes {\rm d}x^{a}, \; \; \; \; E= E^{ia} \tau_{i} \otimes \partial_{a},
\eeq
 where $A^{i}_{a}$, $E^{ia}$ are complex valued functions on $\Sigma$. We fix the standard bilinear complex valued inner product in $ sl(2,{\mbox {\boldmath $C$}}) $ by 
\eq
 k(v,w) := -2{\rm tr}(vw)
\eeq
for any $ v,w \in \sl $. The variables $(A,E)$ are canonically conjugate, the only non-vanishing Poisson bracket is 
\eq
 \{ A^{i}_{a}(x), E^{jb}(y) \} = {\rm i} k^{ij} \delta^{b}_{a} \delta(x,y).
\eeq
 A data $(A,E)$ is accompanied by Lagrange multipliers, a -1 weight density $N$ (the densitized laps function), a vector field $N^{a}$ (the familiar shift) and an $\sl$ valued function $\Lambda$, all defined on $\Sigma$. The Hamiltonian is given by
\eq
 H = \C_{N} + \C_{\vec{N}} + \G_{\Lambda},
\eeq
\eq
 \C_{N} := \int_{\Sigma} {\rm d}^{3}x N \C(A,E) := - \frac{1}{2} \int_{\Sigma} {\rm d}^{3}x N F^{i}_{ab} E^{ja} E^{kb} c_{ijk},
\eeq
\eq
 \C_{\vec{N}} := \int_{\Sigma} {\rm d}^{3}x N^{a} \C_{a}(A,E) := -{\rm i} \int_{\Sigma} {\rm d}^{3}x N^{a} F^{i}_{ab} E^{b}_{i},
\eeq
\eq
 \G_{\Lambda} := \int_{\Sigma} {\rm d}^{3}x \Lambda_{i} \G^{i}(A,E) := {\rm i} \int_{\Sigma} {\rm d}^{3}x \Lambda_{i} D_{a} E^{ia},
\eeq
 where
\eq
 F := \frac{1}{2} F^{i}_{ab} \tau_{i} \otimes {\rm d}x^{a} \wedge {\rm d}x^{b} := {\rm d}A + A \wedge A
\eeq
 is the curvature of $A$, and 
\eq
 D_{a} w^{i} := \partial_{a} w^{i} + c^{i}_{\; jk} A^{j}_{a} w^{k}
\eeq
 is the covariant derivative ($w^{i}$ is a function on $\Sigma$). $ c^{i}_{\; jk}$ are the structure constants of $\sl$ defined by
\eq
 [ \tau_{i} , \tau_{j} ] = c_{\; ij}^{k} \tau_{k} .
\eeq
 The constraints $\C_{N}$, $\C_{\vec{N}}$, $\G_{\Lambda}$ generate respectively the time evolution, diffeomorphisms of $\Sigma$ and the Yang-Mills gauge transformations
\eq
 A \longmapsto g^{-1} A g + g^{-1} {\rm d}g,
\eeq
\eq
 E \longmapsto g^{-1} E g,
\eeq
 where $g$ is any $ SL ( 2, {\mbox {\boldmath $C$}} ) $-valued function on $\Sigma$.

 Apart from the resulting constraint equations, the data $(A,E)$ is subject to the following reality conditions
\eq
 {\rm Im} ( E^{ia} E_{i}^{b} ) = 0 ,
\label{real1}
\eeq
\eq
 {\rm Im} ( \{ E^{ia} E_{i}^{b} , \C_{N} \} ) = 0 .
\label{real2}
\eeq

 As long as the matrix $(E^{ia})_{i,a=1,2,3}$ is of the rank 3 and the signature of the symmetric matrix $(E^{ia}E_{i}^{b})_{a,b=1,2,3}$ is (+,+,+) one constructs an ADM data from $(A,E)$ and the Ashtekar theory is equivalent to the Einstein gravity with 
the Lorentzian signature. However, the theory naturally extends to degenerate cases, when the ranks are lower than 3. 

\subsection*{Classification of degeneracies.}

\hspace{\parindent}
 Since the $E$ field is complex valued, in general the rank of the '2-area matrix' (see e.g. \cite{Lewandowski}) $(E^{ia}E_{i}^{b})$ is lower or equal to the rank of $(E^{ia})$ matrix. If we restrict ourselves to semi-positive definite case of the 2-area 
matrix, the possible cases are (0,0), (1,0), (1,1), (2,1), (2,2) and (3,3), where the numbers indicate the ranks of the triad matrix and the 2-area matrix respectively.

 The examples of triad vector fields falling into specific sectors could be as follows: (0,0) - $E=0$, (1,0) - $E=(\tau_{1}+{\rm i}\tau_{2})\otimes(\frac{\partial}{\partial x^{1}})$, (1,1) - $E=\tau_{1} \otimes (\frac{\partial}{\partial x^{1}})$, (2,1) - $E=(\tau_{1}+{\rm i}\tau_{2})\otimes(\frac{\partial}{\partial x^{1}}) + \tau_{3} \otimes (\frac{\partial}{\partial x^{3}})$, (2,2) - $E=\tau_{1} \otimes (\frac{\partial}{\partial x^{1}}) + \tau_{2} \otimes (\frac{\partial}{\partial x^{2}})$, (3,3) - $E=\tau_{1} \otimes (\frac{\partial}{\partial x^{1}}) + \tau_{2} \otimes (\frac{\partial}{\partial x^{2}}) + \tau_{3} \otimes (\frac{\partial}{\partial x^{3}})$.

\section{Sector (1,0)}

\hspace{\parindent} 
 Sector (1,0) is defined as the one for which ${\rm rank} \left( E^{ia} \right) = 1$, ${\rm sign} \left( E^{ia} E_{i}^{b} \right) = (0,0,0)$ at the surface of initial data $\Sigma$. In this paragraph the Ashtekar equations for the sector (1,0) will be solved. At the beginning, it is useful to choose a convenient gauge. One may show the following
\begin{lemat}
\begin{displaymath}
\begin{array}{c}
 \left[ \left( E^{ia} E_{i}^{b} = 0 \right) \: \wedge \: \left( {\rm rank} \left( E^{ia} \right) = 1 \right) \right] \; \Rightarrow \\
 \Rightarrow \left[ \exists g\in SL(2,{\mbox {\boldmath $C$} } ) \: : \: g^{-1}Eg = \left( \tau_{1} + i\tau_{2} \right) \otimes \left( E^{1a} \partial_{a} \right) \right]
\end{array}
\end{displaymath}
\label{lemat1}
\end{lemat}
 {\bf Proof:} \hspace{\parindent}
 Let us assume that
\eq     
 {\rm rank} (E^{ia}) = 1,
\label{rank1}
\eeq
\eq
 E^{ia} E_{i}^{b} = 0.
\label{rank2}
\eeq
 Equality (\ref{rank1}) implies that 
\eq
 E = \lambda \tau_{1} \otimes E^{3} + \mu \tau_{2} \otimes E^{3} + \tau_{3} \otimes E^{3},
\label{postac}
\eeq
 where $\lambda,\; \mu$ are functions on $ \Sigma $ and $ E^{3} := E^{3a} \partial_{a} \neq 0 $.

 From (\ref{postac}) and (\ref{rank2}) we conclude that 
\eq
 1 + \lambda^{2}  + \mu^{2} = 0.
\label{rownanko}
\eeq
 By the fact from the Appendix we can make such a gauge transformation that $ {\rm Im} \lambda = 0 $.
 It can be easily shown that we can transform $E$ with real $\lambda$ to 
\eq
 E = \lambda^{'} \tau_{1} \otimes E^{3} + \mu \tau_{2} \otimes E^{3},
\eeq
with some new real function $\lambda^{'}$. It can be done by $ g = \left(
\begin{array}{ccc}
 \cos \phi & , & -\sin \phi \\
 \sin \phi & , & \cos \phi 
\end{array}
\right) $ with a suitably chosen $ \phi \in $ {\boldmath $R$} (see Appendix).

 From the Fact 1 it follows that
\eq
 \lambda^{'2} + \mu^{2} = 0 ,
\eeq
 hence $ \mu = \pm {\rm i} \lambda^{'} $.

 Our field variable takes now simple form
\eq
 E = \lambda^{'} ( \tau_{1} \pm {\rm i} \tau_{2} ) \otimes E^{3}.
\label{minus}
\eeq
 By another gauge (with $ g = \left(
\begin{array}{cc}
 {\rm i} & 0 \\
 0 & -{\rm i} 
\end{array}
\right) $ ) we obtain the required form
\eq
 E = ( \tau_{1} + {\rm i} \tau_{2} ) \otimes E^{+},
\eeq
 which ends the proof.

 Now, let us change the basis in $sl(2,{\mbox {\boldmath $C$}})$ to $( \tau_{+}, \tau_{-}, \tau_{0} )$, where $\tau_{+}:=\tau_{1}+i\tau_{2} ,\; \tau_{-}:=\tau_{1}-i\tau_{2} ,\; \tau_{0}:=\tau_{3}$. Expression for the field E takes the simple form
\begin{equation}
 E = \tau_{+} \otimes E^{+} ,
\end{equation}
where $E^{+} := E^{+a} \partial_{a} = E^{1a} \partial_{a}$. It is easy to calculate that in the new basis
\begin{equation}
 c_{+-0}=2{\rm i}=c_{[+-0]},\; {\rm and}
\end{equation}
\begin{equation}
( k_{ij} ) =
\left(
\begin{array}{ccc}
 0 & 2 & 0 \\
 2 & 0 & 0 \\
 0 & 0 & 1 
\end{array}
\right), 
\end{equation}
where $i,j=+,-,0$.

\subsection*{Constraints}

\hspace{\parindent} 
 Constraint equations read now as follows
\begin{displaymath}
 \C \equiv 0, \; \; \G^{-} \equiv 0,
\end{displaymath}
\begin{equation}
\C_{a} = -2{\rm i} \left( i (E^{+}) F^{-} \right)_{a} = 0,
\label{ca}
\end{equation}
\begin{equation}
\G^{0} = -2 i(E^{+}) A^{-} = 0,
\label{g0}
\end{equation}
\begin{equation}
\G^{+} = {\rm i} \partial_{a} E^{+a} + i(E^{+}) A^{0} = 0,
\end{equation}
where $i$ means the inner product and we use the convention for $A^{-}$, $A^{0}$, to be defined analogously to $E^{+}$ and $F^{-}:={\rm d}A^{-}+(A\wedge A)^{-}$. We will use this convention also for the other components of the field variables.
 
 Since $F^{-} = {\rm d}A^{-} - {\rm i}A^{-}\wedge A^{0}$, the following equality is true, provided the constraint equations are fulfilled,
\begin{displaymath}
\begin{array}{c}
 i(E^{+}) \left( {\rm d}A^{-} \wedge A^{-} \right) = i (E^{+}) \left( F^{-} \wedge A^{-} \right) = \\
 =  \left( i(E^{+}) F^{-} \right) \wedge A^{-} + F^{-} \left( i(E^{+}) A^{-} \right) = 0.
\end{array}
\end{displaymath}
Hence the three-form ${\rm d}A^{-} \wedge A^{-} = 0$. Therefore there exist coordinates on $\Sigma$ such that $A^{-} = \alpha {\rm d}\bar{z}$, where $\alpha$ is a function on $\Sigma$ and $\bar{z} = x - {\rm i}y$ ($x,y$ are two of the three real coordinates
 on $\Sigma$) or $\bar{z}\in${\boldmath $R$} (in this case $(x,y,\bar{z})$ are the real coordinates on $\Sigma$).

If $\alpha \neq 0$ we can make gauge transformation with $g={\rm e}^{{\rm i}\lambda \tau_{3}}$, where $\lambda = - \log{\alpha}$. This gives $A^{-} = {\rm d}\bar{z}$ and leaves the form of $E$ unchanged. Indeed, let $g={\rm e}^{\lambda \tau_{0}}$, with $\lambda$ - any complex function on $\Sigma$. We know that $g^{-1}={\rm e}^{-\lambda \tau_{0}}$. Therefore 
\begin{eqnarray*}
 g^{-1} \tau_{\pm} g = {\rm e}^{-\lambda \tau_{0}} \tau_{\pm} {\rm e}^{\lambda \tau_{0}} = {\rm e}^{-\lambda \tau_{0}} \tau_{\pm} (1+ \lambda \tau_{0} + \frac{1}{2} \lambda^{2} \tau_{0}^{2} + \ldots ) = {\rm e}^{-\lambda \tau_{0}} ( \tau_{\pm} + \\
+ \lambda \tau_{\pm} \tau_{0} + \frac{1}{2} \lambda^{2} \tau_{\pm} \tau_{0}^{2} + \ldots ) = {\rm e}^{-\lambda \tau_{0}} ( \tau_{\pm} + \lambda \tau_{0} \tau_{\pm} \mp {\rm i} \lambda \tau_{\pm} + \frac{1}{2} \lambda^{2} \tau_{0} \tau_{\pm} \tau_{0} \mp \\
 \mp \frac{1}{2} {\rm i} \lambda^{2} \tau_{\pm} \tau_{0} + \ldots ) = {\rm e}^{-\lambda \tau_{0}} ( \tau_{\pm} + \lambda \tau_{0} \tau_{\pm} \mp {\rm i} \lambda \tau_{\pm} + \frac{1}{2} \lambda^{2} \tau_{0}^{2} \tau_{\pm} \mp {\rm i} \lambda^{2} \tau_{0} \tau_{\pm} + \\ 
 + \frac{1}{2} ({\rm i}\lambda)^{2} \tau_{\pm} \pm + \ldots ) = {\rm e}^{-\lambda \tau_{0}} {\rm e}^{\lambda (\tau_{0} \mp {\rm i})} \tau_{\pm} = {\rm e}^{\mp \lambda {\rm i}} \tau_{\pm},
\end{eqnarray*}
\begin{displaymath}
 g^{-1}\tau_{0}g=\tau_{0},
\end{displaymath}
\begin{eqnarray*}
 g^{-1}{\rm d}g= {\rm e}^{-\lambda \tau_{0}} {\rm d} ({\rm e}^{\lambda \tau_{0}}) = {\rm e}^{-\lambda \tau_{0}} \left( ({\rm d}\lambda) \tau_{0} + \frac{1}{2} ({\rm d}\lambda^{2}) \tau_{0}^{2} + \ldots \right)= \\
 = {\rm e}^{-\lambda \tau_{0}} ({\rm d}\lambda) \tau_{0} {\rm e}^{\lambda \tau_{0}} = ({\rm d}\lambda) \tau_{0}.
\end{eqnarray*} 

 We will now solve the constraint equations separately for three possible cases.
\begin{enumerate}
\item  $A^{-} = {\rm d}\bar{z}, \; \bar{z} = x - {\rm i}y, \; x,y \in${\boldmath $R$}.\\
 It follows from (\ref{g0}) that
\begin{displaymath}
 E^{+} = E^{+z}\frac{\partial}{\partial z} + E^{+u}\frac{\partial}{\partial u},
\end{displaymath}
where $u\in${\boldmath $R$}, $z=x+{\rm i}y$. Since d$A^{-}=0$, from (\ref{ca}) and (\ref{g0}) we get
\begin{displaymath}
 i (E^{+})  A^{0} = 0 = \G^{+} - {\rm i}\partial_{a}E^{+a},
\end{displaymath}
hence we need to solve the equation 
\begin{equation}
\partial_{a}E^{+a} = 0.
\label{div} 
\end{equation}
The general solution of this equation is 
\begin{equation}
 E^{+a} = \varepsilon^{abc} \Psi_{b,c} ,
\label{Marysia}
\end{equation}
where $\Psi$ is any complex function on $\Sigma$. The condition $E^{+\bar{z}}=0$ gives $\Psi_{z} = \Phi_{,u}$ and $\Psi_{u} = \Phi_{,z}$ with some complex function $\Phi$.

 To solve the constraint equations completely we only have to regard the condition
\begin{displaymath}
 i(E^{+}) A^{0} = 0 .
\end{displaymath}
 This is a simple algebraic equation for $A^{0}$, provided $E^{+}$ is fixed. To end this discussion, it should be noted that there are no constraints for $A^{+}$.
\item $A^{-}={\rm d}\bar{z}, \; \bar{z} \in$ {\boldmath $R$}. \\
 From (\ref{g0}) we get 
\begin{displaymath}
 E^{+} = E^{+x} \frac{\partial}{\partial x} + E^{+y} \frac{\partial}{\partial y}
\end{displaymath}
with $(x,y,\bar{z})$ - coordinates on $\Sigma$.
 It is easy to see that we can solve this case in the same way as we solved point 1. We should only exchange $z$ with $x$ and $u$ with $y$. 
\item $A^{-}=0$.\\ 
 In this case $F^{-}=0$, hence $\C_{a} \equiv 0$. Moreover $\G^{0} \equiv 0$. We only have to solve 
\begin{equation}
 \partial_{a} E^{+a} = {\rm i} E^{+a} A^{0}_{a}.
\label{Ania}
\end{equation}
 For any given $E^{+}$ it is a simple equation for $A^{0}$. We can see that in this case we have no constraints on $E^{+}$ and $A^{+}$.
\end{enumerate}

\subsection*{Evolution equations}
\hspace{\parindent}

 If we take the conditions $ E^{-} = 0, \; E^{0} = 0 $ and $ A^{-} - {\rm d}\bar{z} = 0 $ as the additional constraints, it is easy to see that they weakly commute with the Hamiltonian so their vanishing is preserved by the time evolution provided the constraints are satisfied. In particular the simple form of $E$ is preserved by the time evolution. In fact
\begin{equation}
 \dot{E}^{-a} = -{\rm i} ( c^{-}_{\; \;  -k} E^{-b} + c^{-}_{\; \; 0k} E^{0b} ) ( D_{b} E^{ka} ) = 0, 
\end{equation}
\begin{equation}
 \dot{E}^{0a} = E^{+b} ( \partial_{b} E^{-a} + {\rm i} A^{0}_{b} E^{-a} - {\rm i} A^{-}_{b} E^{0a} ) - E^{-b} D_{b} E^{+a} = 0.
\end{equation}
 The gauge fixing $ A^{-} = {\rm d}\bar{z} $ is also unchanged by the evolution. Namely
\begin{equation}
 \dot{A}^{-}_{a} = E^{-a} F^{0}_{ba} - E^{0a} F^{-}_{ba} = 0.
\end{equation}
 The variable $E$ is independent of time:
\begin{equation}
 \dot{E}^{+a} = E^{+b} ( \partial_{b} E^{0a} + 2{\rm i} A^{-}_{b} E^{+a} - 2{\rm i} A^{+}_{b} E^{-a} ) - E^{0b} D_{b} E^{+a} = 0.
\end{equation}
 Moreover
\begin{equation}
 \dot{A}^{0}_{a} = 2 E^{+b} F^{-}_{ba} - 2 E^{-b} F^{+}_{ba} = 0, \; {\rm and}
\end{equation}
\begin{equation}
 \dot{A}^{+}_{a} = - E^{+b} F^{0}_{ba} + E^{0b} F^{+}_{ba} = E^{+b} ( \partial_{a} A^{0}_{b} - \partial_{b} A^{0}_{a} + 2{\rm i} A^{-}_{a} A^{+}_{b} ).
\label{A+}
\end{equation}
 In order to calculate all the above time derivatives we used constraint equations. We can show that the part of $ A^{+}_{a} $ tangent to $E^{+a}$ is independent of time and the transversal components are linear functions of time. In fact
\begin{displaymath}
 \frac{\partial}{\partial t} ( E^{+a} A^{+}_{a} ) = E^{+a} \dot{A}^{+}_{a} = 2 E^{+a} E^{+b} \partial_{[a} A^{0}_{b]} = 0.
\end{displaymath}
 Hence the right-hand side of (\ref{A+}) is independent of time and $ \frac{\partial}{\partial t} \dot{A}^{+}_{a} = 0 $.
 Now, it can be easily checked that the reality conditions are identically satisfied for the solutions of the constraint and the evolution equations.

\subsection*{Summary}

\hspace{\parindent}
 We have solved completely (1,0) sector of Ashtekar gravity. The general solution for this case (for a certain gauge fixing and choice of coordinates) is as follows. The fields $E^{-}$, $E^{0}$ vanish. Field $E^{+}$ is given by (\ref{Marysia}) and vanishing of the component transversal to $A^{-}$ if $A^{-} \neq 0$ or $E^{+}$ is arbitrary if $A^{-} = 0$. $A^{-}$ is any closed one-form on $\Sigma$, $A^{0}$ is given by the equation (\ref{Ania}) and $A^{+}$ is arbitrary one-form. All the fields are constant in time except of $A^{+}$ which is constant in the direction of $E^{+}$ and is linear in time in the other directions.

 An interesting feature of this solutions is that after imposing certain initial constraints on the field variables at $ t = t_{0} $, at each point they evolve independently from the other points. The points of $ \Sigma $ ``can't see each other during the
 evolution''.

\section{Sector (2,1)}

\hspace{\parindent}
 Sector (2,1) is defined by ${\rm rank} \left( E^{ia} \right) = 2$ and ${\rm sign} \left( E^{ia} E_{i}^{b} \right) = (+,0,0)$ at $t=t_{0}$ (on the surface $\Sigma$). The complete local solution of the Ashtekar-Einstein equations in the sector (2,1) will be given in the present section. We will start from fixing a gauge freedom and a useful choice of coordinates.

\hspace{\parindent}

\begin{lemat}
\begin{displaymath}
\begin{array}{c}
 \left[ \left( {\rm sign} \left( E^{ia} E_{i}^{b} \right) = (+,0,0) \right) \wedge \left( {\rm rank} \left( E^{ia} \right) = 2 \right) \right] \Rightarrow \\ 
\Rightarrow \left[ \exists g \in SL(2,{\mbox {\boldmath $C$} } ) \; : \; g^{-1} E g = \tau_{+} \otimes E^{+} + \tau_{0} \otimes E^{0} \; \; {\rm and} \; \; A^{'0}_{3} = 0 \right], \\
 {\rm where} \; \; A^{'} \stackrel{\rm def}{=} g^{-1} A g + g^{-1} {\rm d}g \; \; {\rm ,and} \; \; \; E^{0} \; {\rm is \; real}.
\end{array}
\end{displaymath}
\label{lemat2}
\end{lemat}
 {\bf Proof:} We assume that 
\eq
 {\rm rank} ( E^{ia} ) = 2,
\label{zal1}
\eeq
\eq
 {\rm sign} ( E^{ia} E_{i}^{b} ) = 1 .
\label{zal2}
\eeq
 Let us choose such a real basis $ (e_{1},e_{2},e_{3}) $ in the tangent space to $\Sigma$ that $ ( E^{ia} E_{i}^{b} ) = {\rm diag} (0,0,1) $. From the fact in the Appendix we conclude that there exists gauge transformation such that 
\eq
 E = E^{kl} \tau_{k} \otimes e_{l} + \tau_{3} \otimes e_{3},
\eeq
 where $ k,l=1,2 $.

 Rank assumption (\ref{zal1}) implies that $ E^{2} = f E^{1} $, where $f$ - complex function on $\Sigma$. (\ref{zal2}) gives $ f = \pm {\rm i} $. Minus sign can be removed in the same way as in (\ref{minus}), which ends the proof.

 From now on let us use the gauge given by the above lemma.
 We can make use of the reality of $ E^{0} $ by choosing convenient coordinate system $ (x^{1}, x^{2}, x^{3} ) $ such that 
\begin{equation}
 E^{0} = \frac{\partial}{\partial x^{3} }.
\label{e0}
\end{equation}

\subsection*{Constraints}
\hspace{\parindent}

 Constraint equations read now as follows
\begin{equation}
 \C = 4{\rm i} E^{+a} E^{0b} F^{-}_{ab} = 0,
\label{sc}
\end{equation}
\begin{equation}
 \C_{a} = E^{0b} F^{0}_{ab} + 2 E^{+b} F^{-}_{ab} = 0,
\label{dif}
\end{equation}
\begin{equation}
 \G^{+} = \partial_{a} E^{+a} + {\rm i} ( A^{+}_{a} E^{0a} - A^{0}_{a} E^{+a} ) = 0,
\label{g1}
\end{equation}
\begin{equation}
 \G^{0} = \partial_{a} E^{0a} + 2{\rm i} A^{-}_{a} E^{+a} = 0,
\label{g2}
\end{equation}
\begin{equation}
 \G^{-} = -{\rm i} A^{-}_{a} E^{0a} = 0.
\label{g3}
\end{equation}
 Due to (\ref{e0}), (\ref{g3}) is solved by $A^{-}_{3}=0$. Since $ \partial_{a} E^{0a} = 0 $, (\ref{g2}) is equivalent to $ A^{-}_{a} E^{+a} = 0 $, or
\begin{equation}
 A^{-}_{1} E^{+1} = - A^{-}_{2} E^{+2}.
\label{row}
\end{equation}
 (\ref{sc}) gives
\begin{displaymath}
 E^{+a} E^{0b} F^{-}_{ab} = - E^{+1} \partial_{3} A^{-}_{1} - E^{+2} \partial_{3} A^{-}_{2} = 0 .
\end{displaymath}
 Let us assume that $ E^{+2} \neq 0 $. Because of (\ref{row}) we have
\begin{equation}
 A^{-}_{1} \partial_{3} A^{-}_{2} = A^{-}_{2} \partial_{3} A^{-}_{1}.
\end{equation}  
 If we assume $ A^{-}_{1} \neq 0 $, this is equivalent to the condition that $ A^{-}_{2} = \Omega A^{-}_{1} $, where $ \Omega $ is a complex function on $ \Sigma $ such that $ \partial_{3} \Omega = 0 $. Thus
\begin{equation}
 A^{-} = A^{-}_{1} \left( {\rm d}x^{1} + \Omega \left( x^{1}, x^{2} \right) {\rm d}x^{2} \right) \; \; {\rm and}
\label{form}
\end{equation}
\begin{equation}
 E^{+} = - \Omega E^{+2} \frac{\partial}{\partial x^{1}} + E^{+2} \frac{\partial}{\partial x^{2}} + E^{+3} \frac{\partial}{\partial x^{3}}.
\label{Kasia}
\end{equation}
 We know, however, that the coordinates $ x^{1}, x^{2} $ can be chosen in such a way that instead of $ \Omega $ we can put i (if $ {\rm Im} \Omega \neq 0 $ ) or $ 0 $ ( if $ {\rm Im} \Omega = 0 $ ). Let us assume then, that from now on $ \Omega = 0 $ or $
\Omega = $i.

 In order to solve constraints completely we have to solve two more equations, namely (\ref{dif}) and (\ref{g1}). Straightforward calculation shows that 
\begin{displaymath}
 E^{+b} F^{-}_{ab} = - E^{+b} \partial_{b} A^{-}_{a} - {\rm i} E^{+b} A^{0}_{b} A^{-}_{a}, \; \; {\rm and}
\end{displaymath}
\begin{displaymath}
 E^{0b} F^{0}_{ab} = - \partial_{3} A^{0}_{a} + 2{\rm i} A^{-}_{a} A^{+}_{3} .
\end{displaymath} 
 Hence (\ref{dif}) gives
\begin{displaymath}
 2 E^{+b} \partial_{b} A^{-}_{a} + \partial_{3} A^{0}_{a} = 2{\rm i} A^{-}_{a} ( A^{+}_{3} - E^{+b} A^{0}_{b} ) .
\end{displaymath}
 Substituting (\ref{g1}) into the above equation gives
\begin{equation}
 \partial_{3} A^{0}_{a} = -2 \partial_{b} ( E^{+b} A^{-}_{a} ).
\label{Patrycja}
\end{equation}
 With a given $ E^{+} $ and $ A^{-} $, the above equation describes the dependence of $ A^{0} $ on the coordinate $ x^{3} $.

 To end the analysis of the constraints we should add (\ref{g1}), which can be treated as the constraint on $ A^{+}_{3} $, provided $ E^{+}, \; A^{0} $ are known
\begin{equation}
 A^{+}_{3} = {\rm i} \partial_{a} E^{+a} + A^{0}_{a} E^{+a}.
\label{talk}
\end{equation}
 At last, the case $ E^{+1} = E^{+2} = 0 $ should be considered separately. However, the only difference in the family of solutions for this case is in the form of $ A^{-} $. Now, we have no restrictions on $ A^{-}_{1} $ and $ A^{-}_{2} $.

 For $ A^{-}_{1} = 0 $ we get from (\ref{row}) that $ A^{-}_{2} = 0 $ or $ E^{+2} = 0 $, but these cases are included in the other ones.

 Hence we have solved completely constraint equations for the sector (2,1).

\subsection*{Evolution}
\hspace{\parindent}

 Let us now consider conditions $ E^{-} = 0, \; E^{0} - \frac{\partial}{\partial x^{3}} = 0 , \; A^{-}_{3} = 0 $ as the new additional constraints on the initial data. One can show that they weakly commute with the Hamiltonian, hence they are preserved by the evolution. In fact
\eq
 \dot{E}^{-a} = E^{0b} \partial_{b} E^{-a} + {\rm i} \G^{-} E^{0a} + {\rm i} E^{0b} A^{0}_{b} E^{-a} - E^{-b} D_{b} E^{0a} = 0,
\eeq
\eq
 \dot{E}^{0a} = -2 E^{+b} \partial_{b} E^{-a} - E^{0a} \partial_{b} E^{0b} + \G^{0} E^{0a} - 2{\rm i} E^{+b} A^{0}_{b} E^{-a} + 2 E^{-b} D_{b} E^{+a} = 0,
\eeq
\eq
 \dot{A}^{0}_{3} = 2 E^{+b} F^{-}_{b3} - 2 E^{-b} F^{+}_{b3} = E^{0b} F^{0}_{3b} = 0.
\label{a03}
\eeq
 Moreover, due to constraint equations, we get
\eq
 \dot{A}^{-}_{a} = E^{-b} F^{0}_{ba} - E^{0b} F^{-}_{ba} = F^{-}_{a3}, \; \; {\rm thus}
\eeq
\eq
 \dot{A}^{-}_{3} = 0,
\eeq
\eq
 \dot{A}^{-}_{1} = - \partial_{3} A^{-}_{1},
\label{A1}
\eeq
\eq
 \dot{A}^{-}_{2} = - \partial_{3} A^{-}_{2}.
\label{A2}
\eeq    
 In order to find the evolution of $ E^{+} $, let us first calculate 
\eq
 \dot{A}^{+}_{3} = - E^{+b} F^{0}_{b3} + E^{0b} F^{+}_{b3} = - E^{0a} E^{+b} F^{0}_{ba} = - E^{+b} \C_{b} = 0.
\eeq
 Now we have
\eq
 \dot{E}^{+a} = -{\rm i} c^{+}_{\; \; ij} E^{ib} ( \partial_{b} E^{ja} + c^{j}_{\; \; kl} A^{k}_{b} E^{la} ), \; {\rm thus}
\eeq
\begin{eqnarray*}
 \dot{E}^{+a} = E^{+b} \partial_{b} E^{0a} - E^{0b} \partial_{b} E^{+a} - 2{\rm i} A^{+}_{b} E^{-b} E^{+a} + E^{+a} \G^{0} - \\ 
2{\rm i} (\partial_{b} E^{0b} ) E^{+a} - {\rm i} E^{0a} E^{0b} A^{+}_{b} + {\rm i} E^{+a} E^{0b} A^{0}_{b}, 
\end{eqnarray*}

\noindent and since the constraints show that $ E^{0b} A^{0}_{b} = A^{0}_{3} = 0 $, we get
\eq
 \dot{E}^{+a} = - \partial_{3} E^{+a} - {\rm i} E^{0a} A^{+}_{3}.
\label{e+a}
\eeq
 Since $ E^{0a} A^{+}_{3} $ does not depend on time, (\ref{e+a}) can be easily integrated for $ E^{+a}(t) $.

 We obtain similar equations for the components of $A^{0}$. In the same way as in (\ref{a03}) we get that $ \dot{A}^{0}_{a} = F^{0}_{a3} $, hence
\eq
 \dot{A}^{0}_{a} = - \partial_{3} A^{0}_{a} + 2{\rm i} A^{-}_{a} A^{+}_{3}.
\label{Beata}
\eeq
 Again we have simple linear equation for $A^{0}$.

 The last thing we need to do to solve completely the evolution equations is to find the function $A^{+}(t)$. Let us calculate
\eq
 \dot{A}^{+}_{a} = - E^{+b} F^{0}_{ba} + E^{0b} F^{+}_{ba}.
\eeq
 This gives
\begin{eqnarray*}
 \dot{A}^{+}_{a} = -2 E^{+b} \partial_{[b} A^{0}_{a]} - \G^{0} A^{+}_{a} + ( \partial_{b} E^{0b} ) A^{+}_{a} + \\
 +2{\rm i} E^{+b} A^{+}_{b} A^{-}_{a} + 2 E^{0b} \partial_{[b} A^{+}_{a]} + 2{\rm i} E^{0b} A^{+}_{[b} A^{0}_{a]}.
\end{eqnarray*}

\noindent Using the constraints we get
\eq
 \dot{A}^{+}_{a} = -2 E^{+b} \partial_{[b} A^{0}_{a]} + 2{\rm i} E^{+b} A^{+}_{b} A^{-}_{a} + 2 \partial_{[3} A^{+}_{a]} + 2{\rm i} A^{+}_{[3} A^{0}_{a]}, \; {\rm thus}
\eeq
 \eq
\dot{A}^{+}_{a} = \partial_{3} A^{+}_{a} + 2{\rm i} A^{-}_{a} E^{+b} A^{+}_{b} - \partial_{a} A^{+}_{3} + {\rm i} A^{0}_{a} A^{+}_{3} - 2 E^{+b} \partial_{[b} A^{0}_{a]}.
\label{a+a}
\eeq
 We can see that due to the second term on the right-hand side of the above equation $A^{+}_{1}$ depends on $A^{+}_{2}$ and conversely. However, we can simplify this equation using the results of constraint analysis. Let us consider two different possibilities.
\begin{enumerate}
\item $ E^{+1} = E^{+2} = 0 $. \\
 In this case $ E^{+b} A^{+}_{b} = 0 $ and we get simple linear equations for $A^{+}_{1}$ and $A^{+}_{2}$, namely
\eq
 \dot{A}^{+}_{a} = - \partial_{3} A^{+}_{a} - \partial_{a} A^{+}_{3} + {\rm i} A^{0}_{a} A^{+}_{3} - 2 E^{+b} \partial_{[b} A^{0}_{a]}.
\eeq
\item $ E^{+2} \neq 0, \; E^{+1} = - \Omega E^{+2} $ ( $\Omega=0$ or $\Omega=$i ). \\
 It is easy to calculate that 
\begin{displaymath}
 \frac{\partial}{\partial t} ( A^{+}_{2} - \Omega A^{+}_{1} ) = \partial_{3} ( A^{+}_{2} - \Omega A^{+}_{1} ) - ( \partial_{2} - {\Omega}\partial_{1} ) A^{+}_{3} + {\rm i}A^{+}_{3} ( A^{0}_{2} - {\Omega}A^{0}_{1} ) -
\end{displaymath}
$ - E^{+b} \left[ \partial_{b} \left( A^{0}_{2} - {\Omega} A^{0}_{1} \right) - \left( \partial_{2} - {\Omega}\partial_{1} \right) A^{0}_{b} \right].$
\bigskip

 Hence we have a simple linear equation for $ ( A^{+}_{2} - \Omega A^{+}_{1} ) ( t ) $. Substituting $ \Omega A^{+}_{1}(t) + ( A^{+}_{2} - \Omega A^{+}_{1} ) ( t ) $ for $ A^{+}_{2}(t) $ in (\ref{a+a}) we get the linear equation for $ A^{+}_{1}(t) $. It can be integrated if $ A^{-}, \; E^{+}, \; A^{+}_{3}, \; A^{0} $ are known.
\end{enumerate}

 This solves the evolution equations. One can see that the reality conditions are satisfied for all the solutions we have found.

\subsection*{Summary}

\hspace{\parindent}
 Let us summarize the general solution of the Ashtekar-Einstein equations in the sector (2,1). First, we have $E^{-}(t)=0$ and $E^{0}(t)=\frac{\partial}{\partial x^{3}}$. The fields $E^{+a}$ propagate along the integral curves of $E^{0}$ according to the equation (\ref{e+a}). The components $E^{+2}$ and $E^{+3}$ are arbitrary functions of the `spatial` coordinates (but $E^{+3} \neq 0$) and the remaining component is given by $ E^{+1} = -\Omega E^{+2} $, equation (\ref{Kasia}) ($\Omega=0$ or $\Omega=$i). If $E^{+2} \neq 0$, $A^{-}$ is given by (\ref{form}) with the same $\Omega$ as above, and if $E^{+2} = 0$, arbitrary one-form $A^{-}$ with $A^{-}_{3}=0$ is the solution. Fields $A^{-}_{1}$ and $A^{-}_{2}$ propagate along the integral curves of $E^{0}$ at the speed of light. $A^{0}$ is any field which propagates along the same curves as $A^{-}$ and $E^{+}$ according to equation (\ref{Beata}) and depends on the coordinate $x^{3}$ according to the equation (\ref{Patrycja}).

 The field $A^{+}_{3}$ does not depend on time and is given by the equation (\ref{talk}). $A^{+}_{1}$ and $A^{+}_{2}$ are any functions on $\Sigma$ with the dependence on time given by (\ref{a+a}).  

 It should be noted that, as in sector (2,2), the characteristic feature of our solutions is the fact that evolution takes place on the curves, namely curves defined by $x^{1},x^{2}=const$. During the evolution these curves do not interact. 

\section{Concluding Remarks}

\hspace{\parindent} 
 As indicated in the introduction, all the possible degenerate sectors of Ashtekar's gravity have been solved. They all have certain important features in common.

 First of all, the conditions defining the degeneracy sectors weakly commute with the Hamiltonian. Therefore, if $t=t_{0}$ corresponds to the surface of initial data $\Sigma$, then there is an $\varepsilon>0$ such that for all $t$ between $t_{0}$ and $t_{0}+\varepsilon$ degeneracy type is the same (evolution preserves the degeneracy locally, where the word ``local'' refers to both space and time). Hence if the initial data on $\Sigma$ is specified in such a way that all of it belongs to the same degeneracy sector, the generic behavior will be such that the evolution preserves the character of the degeneracy. On the other hand, if there are regions on $\Sigma$ with different types of data then the above need not be true (see \cite{recent}).

 The other important feature is that for all the sectors, the surface of initial data $\Sigma$ is foliated by sub-manifolds of the dimension equal to the rank of the densitized inverse three-metric $qq^{ab}$ on $\Sigma$. The evolution always takes place in such a way that these sub-manifolds evolve independently. The time derivatives of the field variables on a fixed leaf of the foliation depend only on the values of these fields on the leaf and on the derivatives along the leaf. $qq^{ab}$ decides that the fields evolve  along the surfaces \cite{Lewandowski}, along the curves (sector (2,1) and \cite{Jacobson}), at the points independently (sector (1,0)) or do not evolve at all ($E^{ia}=0$ for all $i,a$).

\subsection*{Acknowledgments}

\hspace{\parindent}
J.L. was supported by Alexander von Humboldt-Stiftung and the Polish
Committee on Scientific Research (KBN, grant no. 2 P03B 017 12).
J.W. was supported by Polish Ministry of Education and Stefan Batory Trust.

\hspace{\parindent}

\newpage

\section{Appendix}

\appendix

\subsection*{Useful Fact}

\begin{fakt}
 Let $u^{i}$, $v^{i}$, $w^{i}$ be such vectors in $\sl$ that $ u^{i} u_{i} = v^{i} v_{i} = w^{i} w_{i} = 1 $ and $ u^{i} v_{i} = u^{i} w_{i} = v^{i} w_{i} = 0 $. Then, there exists $ g \in SL ( 2, {\mbox {\boldmath $C$}} ) $ such that 
\begin{displaymath}
 g^{-1} u g = \tau_{1}, \; \; \; g^{-1} v g = \tau_{2}, \; \; \; g^{-1} w g = \tau_{3},
\end{displaymath}
 where $(\tau_{i})$ is an orthonormal basis in $\sl$ such that $ [ \tau_{i} , \tau_{j} ] = \varepsilon_{ij}^{\; \; \; k} \tau_{k} $.
\end{fakt}
 {\bf Proof}: Let us fix
\eq
 \tau_{1} = \frac{1}{2} \left(
\begin{array}{cc}
 0 & {\rm i} \\
 {\rm i} & 0 
\end{array}
\right), \; \; \; 
\tau_{2} = \frac{1}{2} \left( 
\begin{array}{cc}
 {\rm i} & 0 \\
 0 & -{\rm i}
\end{array}
\right), \; \; \;
\tau_{3} = 
\frac{1}{2} \left(
\begin{array}{cc}
 0 & 1 \\
 -1 & 0 
\end{array} 
\right).
\eeq
 Let us check what transformation is made by 
\eq
 g_{3} = \left(
\begin{array}{cc}
 \cos \phi & {\rm i} \sin \phi \\
 {\rm i} \sin \phi & \cos \phi
\end{array}
\right) \in SL ( 2, {\mbox {\boldmath $C$}} ),
\eeq
with $ \phi \in ${\boldmath $R$}. It is easy to calculate that
\eq
 g^{-1}_{3} \tau_{1} g_{3} = \cos (2\phi) \tau_{1} - \sin (2\phi) \tau_{2},
\eeq
\eq
 g^{-1}_{3} \tau_{2} g_{3} = \sin (2\phi) \tau_{1} + \cos (2\phi) \tau_{2},
\eeq
\eq
 g^{-1}_{3} \tau_{3} g_{3} = \tau_{3}.
\eeq
 Hence, choosing proper $\phi \in ${\boldmath $R$} we can make any rotation around the $\tau_{3}$-axis in the vector space $\sl$.

 Analogously we can check that
\eq
 g_{1} = \left(
\begin{array}{cc}
 {\rm e}^{{\rm i}\phi} & 0 \\
 0 & {\rm e}^{-{\rm i}\phi}
\end{array}
\right) \; \; \; {\rm and} \; \; \; g_{2} = \left( 
\begin{array}{cc}
 \cos \phi & - \sin \phi \\
 \sin \phi & \cos \phi 
\end{array}
\right)
\eeq
give rotations around the $\tau_{1}$-, $\tau_{2}$-axes respectively. 

 The above fact follows now from the properties of rotations in three-dimensional vector space.
 
\subsection*{Complete set of local solutions for degenerate gravity.}

\hspace{\parindent}
 We shall list here the general solutions obtained for all possible kinds of degeneracy which can potentially occur in Ashtekar's theory for the Lorentzian signature. The interpretation and some properties of these solutions are given in the preceding sections.\\

{\bf Sector (0,0)}\\

 $E^{ia}=0$, \hspace{20pt} $A^{i}_{a}$ - arbitrary, constant in time.\\

{\bf Sector(1,0)}\\

 $E^{-a}=0$, \hspace{15pt} $E^{0a}=0$.\\

 There are two possibilities for $E^{+}$ and $A^{-}$ :

 1. $A^{-}_{a}=0$, $E^{+a}$ - arbitrary, constant in time, or
 
 2. $A^{-}_{a}=({\rm d} \bar{z})_{a}$ ($\bar{z}$ is a real or complex coordinate on $\Sigma$), $E^{+a}= \varepsilon^{abc} \Psi_{b,c}$, where $\Psi_{a}$ is constant in time and $\Psi_ {z}=\Phi_{,u}$, $\Psi_{u}=\Phi_{,z}$ with $\Phi$ - an arbitrary function
 on $\Sigma$ and $z_{,\bar{z}}=0$, $u_{,\bar{z}}=0$.\\

 $A^{0}_{a}$ given by: $\partial_{a} E^{+a} = {\rm i} E^{+a} A^{0}_{a}$,\\

 $A^{+}_{a}$ given by: $\dot{A}^{+}_{a} =  E^{+b} ( \partial_{a} A^{0}_{b} - \partial_{b} A^{0}_{a} + 2{\rm i} A^{-}_{a} A^{+}_{b} )$.\\

{\bf Sector(1,1)}\\

 $E^{1a}=E^{2a}=0$, \hspace{15pt} $E^{3a}=(\frac{\partial}{\partial{x^{3}}})^{a}$, \hspace{15pt} $A^{i}_{3}=0$, \hspace{15pt} $A^{3}_{a}$ - arbitrary, constant in time and $x^{3}$,\\ 

 $A^{1}_{a}$, $A^{2}_{a}$ given by: $\partial_{t}(A^{1}_{a} \pm {\rm i} A^{2}_{a})= \pm \partial_{3}(A^{1}_{a} \pm {\rm i} A^{2}_{a})$.
\\

{\bf Sector(2,1)}\\

  $E^{-a}=0$, \hspace{15pt} $E^{0a}=(\frac{\partial}{\partial x^{3}})^{a}$, \hspace{15pt} ($E^{+2}$, $E^{+3}$) - arbitrary functions of spatial coordinates ($E^{+3} \neq 0$),

  $E^{+1} = - \Omega E^{+2}$ ($\Omega = 0$ or $\Omega = {\rm i}$). \\

 If $E^{+2} \neq 0$, then $A^{-}_{a} = A^{-}_{1} \left( {\rm d}x^{1} + \Omega {\rm d}x^{2} \right)_{a}$. 

 If $E^{+2}=0$, then $A^{-}_{a}$ - arbitrary function of spatial coordinates with $A^{-}_{3}=0$.\\ 

 For both cases:  $\dot{A}^{-}_{1} = - \partial_{3} A^{-}_{1}$ and $\dot{A}^{-}_{2} = - \partial_{3} A^{-}_{2}$.\\ 

 $A^{0}_{a}$ depends on spatial coordinates according to: 

 $\partial_{3} A^{0}_{a} = -2 \partial_{b} ( E^{+b} A^{-}_{a} )$.\\ 

 $A^{+}_{3}$ is constant in time and given by: 

 $A^{+}_{3} = {\rm i} \partial_{a} E^{+a} + A^{0}_{a} E^{+a}$.\\ 

 Evolution of $A^{0}_{a}$ is given by: 

 $\dot{A}^{0}_{a} = - \partial_{3} A^{0}_{a} + 2{\rm i} A^{-}_{a} A^{+}_{3} $.\\ 

 Evolution of $E^{+a}$ is determined from: 

 $\dot{E}^{+a} = - \partial_{3} E^{+a} - {\rm i} E^{0a} A^{+}_{3}$.\\ 

 $A^{+}_{1}$ and $A^{+}_{2}$ are arbitrary functions evolving according to:\\

 1. $\dot{A}^{+}_{a} = - \partial_{3} A^{+}_{a} - \partial_{a} A^{+}_{3} + {\rm i} A^{0}_{a} A^{+}_{3} - 2 E^{+b} \partial_{[b} A^{0}_{a]}$, if $E^{+1}=E^{+2}=0$, or\\

 2. $ \frac{\partial}{\partial t} ( A^{+}_{2} \pm \Omega A^{+}_{1} ) = \partial_{3} ( A^{+}_{2} \pm \Omega A^{+}_{1} ) - ( \partial_{2} \pm {\Omega}\partial_{1} ) A^{+}_{3} + {\rm i}A^{+}_{3} ( A^{0}_{2} \pm {\Omega}A^{0}_{1} ) -$

$ - E^{+b} \left[ \partial_{b} \left( A^{0}_{2} \pm {\Omega} A^{0}_{1} \right) - \left( \partial_{2} \pm {\Omega}\partial_{1} \right) A^{0}_{b} \right]$, if $E^{+2} \neq 0$.\\

{\bf Sector(2,2)}\\

 $E^{1a}=(\frac{\partial}{\partial x^{1}})^{a}$, \hspace{15pt} $E^{2a}=(\frac{\partial}{\partial x^{2}})^{a}$, \hspace{15pt} $E^{3a}=0$, 

 $A^{i}_{1}=0$, \hspace{15pt} $A^{i}_{2}=0$, \hspace{15pt} $A^{1}_{3}=-\lambda_{,2}$, \hspace{15pt} $A^{2}_{3}=\lambda_{,1}$, \hspace{15pt} $A^{3}_{3}={\rm i}\lambda_{,t}$,\\

  with $\lambda$ - any complex function satisfying: 

 $\lambda_{,tt} - \lambda_{,11} - \lambda_{,22} = 0$.


\newpage

\begin{thebibliography}{69}
\bibitem{Ashtekar} A. Ashtekar, {\em Lectures on Non-perturbative Canonical Gravity} (World Scientific, 1991).
\bibitem{Bengtsson} I. Bengtsson, ``Some Observations on Degenerate Metrics'', {\em Gen. Rel. Grav.} {\bf 25} (1993) 101, ``A new phase for general relativity?'', {\em Class. Quantum Grav.} {\bf 7} (1990) 27
\bibitem{Romano} T. Jacobson and J.D. Romano, ``Degenerate extensions of general relativity'', {\em Class. Quantum Grav.} {\bf 9} (1992) L119
\bibitem{Reisenberger} M. P. Reisenberger, ``New constraints for canonical general relativity'', {\em Nucl. Phys.} {\bf B} 457 (1995) 643-687
\bibitem{Matschull} H.J. Matschull, ``Causal Structure and Diffeomorphisms in Ashtekar's Gravity'', {\em Class. Quantum Grav.} {\bf 13} (1996) 765-782.
\bibitem{Jacobson} T. Jacobson, ``1+1 Sector of 3+1 Gravity'', {\em Class. Quantum Grav.} {\bf 13} (1996) L111-L116.
\bibitem{Lewandowski} J. Lewandowski and J. Wi\'sniewski, ``2+1 Sector of 3+1 Gravity'', {\em Class. Quantum Grav.} {\bf 14} (1997) 775-782.
\bibitem{Ma} Y. Ma and C. Liang, ``The Degenerate Sector of Ashtekar's Phase Space'', {\em Modern Phys. Letters A}, Vol. 13, No. 35 (1998) 2839-2843 
\bibitem{recent} I. Bengtsson and T. Jacobson, ``Degenerate metric phase boundaries'', {\em Class. Quantum Grav.} {\bf 14} (1997) 3109
\end{thebibliography}
\end{document}